# A Vibe Coding Learning Design To Enhance EFL Students' Talking To, Through, and About AI

David James Woo [a], Kai Guo [b, *], and Yangyang Yu [c]

[a] Everwrite Limited, Hong Kong, China
[b] The University of Hong Kong, Hong Kong, China
[c] Shanghai Jiao Tong University, Shanghai, China

[*] **Corresponding author**
  - Email address: kaiguo@connect.hku.hk
  - Postal address: Room 901, T.T. Tsui Building, The University of Hong Kong, Pokfulam Road, Hong Kong, China

**FUNDING DETAILS**

This research did not receive any specific grant from funding agencies in the public, commercial, or not-for-profit sectors.

**DECLARATION OF CONFLICTING INTERESTS**

The authors report there are no competing interests to declare.

**Declaration of Generative AI and AI-assisted technologies in the writing process**

During the preparation of this work the authors used ChatGPT in order to improve readability and language. After using this tool, the authors reviewed and edited the content as needed and take full responsibility for the content of the publication.

# A Vibe Coding Learning Design To Enhance EFL Students' Talking To, Through, and About AI

**Abstract**
This Innovative Practice article reports on piloting of vibe coding -- using natural language to create software applications with AI -- for English as a foreign language (EFL) education. We developed a human-AI metalanguaging framework with three dimensions: *talking to AI* (prompt engineering), *talking through AI* (negotiating authorship), and *talking about AI* (mental models of AI). Using backward design principles, we created a four-hour workshop where two students designed applications addressing authentic EFL writing challenges. We adopted a case study methodology collecting data from worksheets, video recordings, think-aloud protocols, screen recordings, and AI-generated images. Contrasting cases showed one student successfully vibe coding a functional application cohering to her intended design while another encountered technical difficulties with major gaps between intended design and actual functionality. Analysis reveals differences in students' prompt engineering approaches, suggesting different AI mental models and tensions in attributing authorship. We argue that AI functions as a beneficial languaging machine and that differences in how students talk to, through, and about AI explain vibe coding outcome variations. Findings indicate effective vibe coding instruction requires explicit metalanguaging scaffolding: teaching structured prompt engineering, facilitating critical authorship discussions, and developing vocabulary for articulating AI mental models.

**Keywords**
AI literacy; design thinking; vibe coding; EFL writing; prompt engineering; student agency

## 1. Introduction

Artificial intelligence (AI) applications for English as a foreign language (EFL) have expanded as teachers and students use large language models (LLMs) with chatbot interfaces like ChatGPT to generate language outputs that support learning (Barrot, 2023). Significantly, LLMs have mastered human and computer languages and can be equipped with tools to perform tasks autonomously. As a result, people without coding knowledge are increasingly capable of developing software through *vibe coding*, which refers to people producing software through natural language interactions with LLMs. Vibe coding extends language learning beyond personalizing chatbot writing assistants (Guo & Li, 2024), enabling EFL students to design a variety of applications (interactive games, mobile apps) that address specific writing challenges.

Vibe coding directly impacts EFL students' language learning by turning their natural language communication with LLMs into an act of embodied learning (Vygotsky, 1978). Building on recent developments in AI-mediated language education (Jones, 2025), we conceptualize how EFL students interact with LLMs while vibe coding as *talking to AI, talking through AI*, and *talking about AI*. *Talking to AI* refers to prompt engineering, that is, crafting appropriate instructions to an LLM so that it generates what the user wants. To do so requires metalinguistic and metacognitive skills as an EFL student adapts communication strategies for an unfamiliar actor (Woo et al., 2024). *Talking through AI* refers to negotiating authorship and voice in completing a human-AI composition, influencing whether the composition could be considered human-authored or synthetic (Knowles, 2024). *Talking about AI* refers to EFL students' mental models of AI (Pataranutaporn et al., 2023), or how students imagine AI. As illustrated in Figure 1, these three dimensions are interrelated. For example, students' mental models of AI can affect their interaction with AI, which in turn influence their perception of AI, and possibly how they negotiate the authorship while working with machine-in-the-loop (Pataranutaporn et al., 2023).

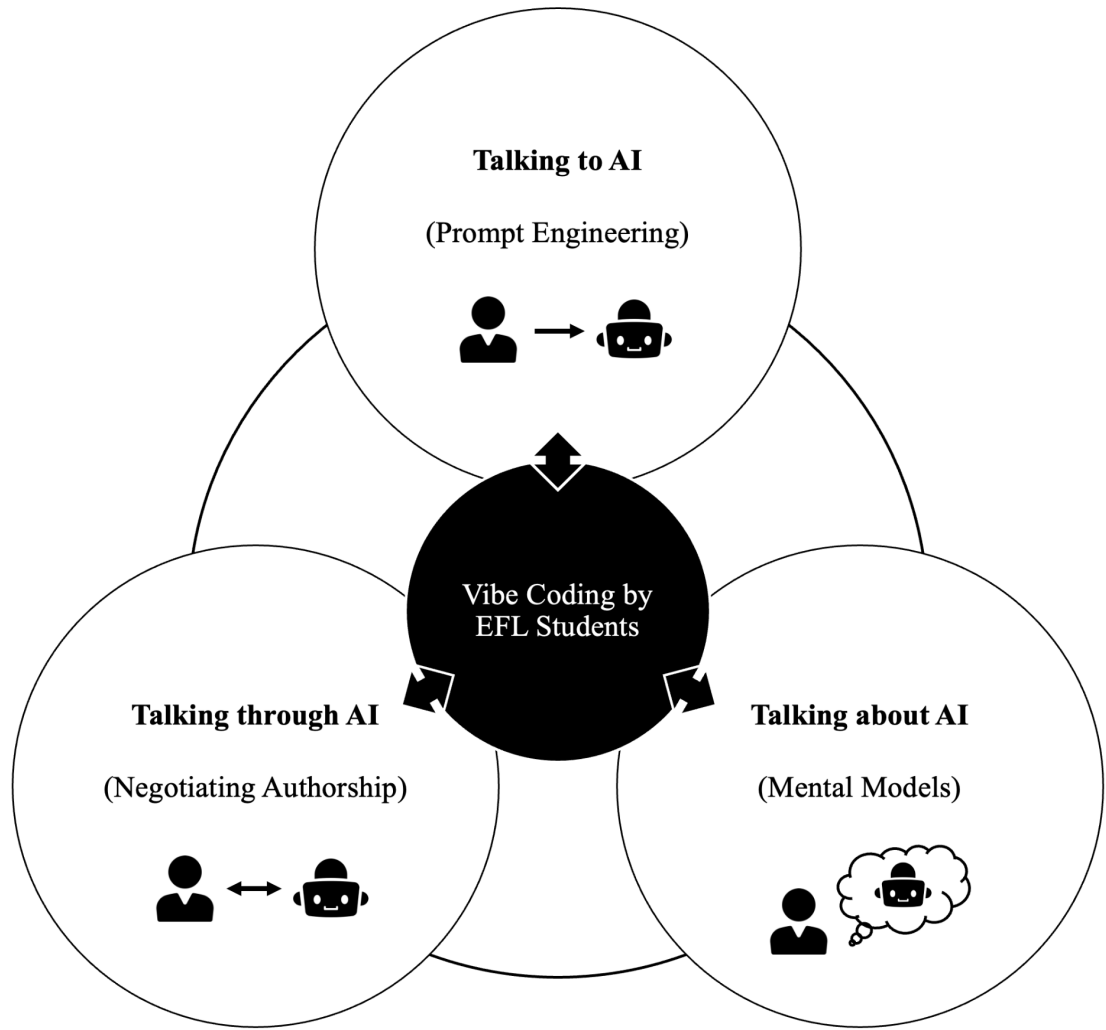

**Figure 1**. Talking to, through, and about AI
**Alt text**: A framework presenting how EFL students interact with AI through three modes: talking to AI (prompting engineering), through AI (negotiating authorship), and about AI (building mental models)

We frame vibe coding through these three dimensions and conceptualize AI as a beneficial languaging machine: AI provokes not only students' language but more importantly their reflection and self-regulation. These impact students' AI literacies as well as their agency, creativity, and engagement. However, vibe coding education for EFL contexts is untested. Little is known about how students interact with AI—whether by talking to it, through it, or about it—and how these interactions shape the three dimensions of AI-mediated language education. This Innovative Practice article explores EFL students' learning process from the implementation of vibe coding education, comparing cases to inform teachers' effective vibe coding instruction.

## 2. Materials and Methods

### *2.1. Context and Participants*

We developed the innovative practice as a pilot for an annual human–AI creative writing contest delivering AI and English literacy workshops in Hong Kong secondary schools. The first author piloted the 2025-26 contest learning design in a 4-hour English-medium workshop on July 17, 2025, at an all-girls Hong Kong secondary school. It was expected that after the pilot implementation and evaluation, the revised learning design would be implemented in participating schools.

Two Form Three students (US Grade Level 9, ages 14-15), Student A and Student G volunteered; both were strong EFL performers in the same class taught by the first author. They were informed of the study and their rights as participants. Prior to the workshop, both students reported having worked on design projects, and having used AI chatbots.

*2.2. Learning Designs*

We developed the learning design from scratch given vibe coding's novelty. We approached development with a backward design (Wiggins & McTighe, 1998), considering first what learners should understand and be able to do: the long-term performance goal is that students should be able to vibe code software applications as solutions to address authentic language learning problem scenarios. To do so, students should understand that their language capabilities impact and are impacted by vibe coding. Students should also understand design thinking (DT), an iterative problem-solving process (Razzouk & Shute, 2012). Applied to EFL contexts, DT could enhance students' language skills, including speaking (Buphate & Esteban, 2022) and writing (Wible, 2020). DT serves as a guiding framework for students to approach vibe coding.

We specified essential understandings through intended learning outcomes (ILOs). For DT ILOs, we adopted those from the Middle Years Program (MYP) design cycle because it emphasizes creative, critical thinking for real-world problems, and a coherent four-stage structure (Inquiring and Analyzing; Developing Ideas; Creating the Solution; Evaluating) (International Baccalaureate Organization, 2025). For vibe-coding ILOs, we selected ILOs from the affective and cognitive learning dimensions of Ng et al.'s (2023) framework.

We determined acceptable evidence for our ILOs. Given our learning design's four-hour time scope, we adopted informal checks for understanding along with observation and dialogue, and prompts. Importantly, we implemented a performance task so we might practically assess students' ability to vibe code software applications as solutions to address authentic language learning problem scenarios.

We sequenced teaching and learning activities. First, the teacher provided facilitative teaching and direct instruction on DT. Students explored the MYP design cycle and process items through thinking routines (e.g., See-Think-Wonder, Think-Pair-Share, ZoomIn-ZoomOut, Observation and Explanation, Headline and Stop and Question) (Harvard Graduate School of Education, 2022). They then developed conceptual understanding of vibe coding concepts (generative AI, LLMs, prompt engineering) through additional routines (e.g., List-Group-Label, Hear-Think-Wonder). We created a slide deck to introduce these concepts and activities. Students wrote notes and provided verbal responses during these activities. We video recorded students' notes and verbal responses.

The summative performance task (see Figure 2) was an authentic problem scenario emphasizing EFL writing. The teacher guided students through the MYP design cycle to inquire and analyze, and develop ideas. In that way, we had composed worksheets with prompts for different design cycle stages. Students completed the prompts individually and the teacher also completed the prompts as a model. The students wrote, drew diagrams and sketches and completed information tables. Students and teachers verbally reported their progress to each other. We video recorded students' writing and verbal reports.

> You are an English language teacher in a Hong Kong secondary school. You are passionate about enhancing students' English language education through vibe-coding generative artificial intelligence (AI) applications.
>
> Form 4 students want to practice process writing independently. Specifically, they want to practice planning, drafting and reviewing HKDSE English Language Paper 2 Writing compositions. Design an app that can score HKDSE English compositions according to the HKDSE marking guidelines and provide feedback on how to improve like your English teacher.
>
> Vibe code your app from any generative AI software. Your app must be accessible to others from the Internet.

**Figure 2.** The authentic problem scenario
**Alt text**: Three paragraphs comprising a writing task prompt

Students then iteratively created and evaluated their solution. Students chose their own vibe coding software and vibe coded their application on their iPads. We provided students with soft copies of 1) the HKDSE marking guidelines and 2) sample HKDSE English compositions with scores and feedback. We observed students as they vibe coded and also recorded their screens as evidence. For students to verbally evaluate their progress, we implemented and video recorded a think-aloud protocol. The protocol's design was that every five minutes, the first author asked students six questions and students were free to respond to any extent: 1) How complete is your app from 0 to 100%? 2) What percentage of the app's ideas are your ideas? 3) What percentage of the app's ideas are AI's? 4) What do you think about this prompt? 5) What do you think about this output? 6. How do you feel?

Near the end of the workshop, students submitted their apps for our evaluation on a Google Form. Students then completed final evaluation prompts in their worksheets and verbally reported their responses, which was video recorded. At the workshop's conclusion, we collected all student worksheets (see Supplemental Materials) and observational notes, and transcribed the video recordings.

## 3. Results

We explored how students talk to, through, and about AI while vibe coding, because those languaging dimensions impact and are impacted by students' vibe coding software applications as solutions to address authentic language learning problem scenarios. We developed understanding by case study, since we had two students and collected much data from them.

We performed a content analysis on the data we had collected. To address how students talk to AI, we transcribed students' prompts from their screen recordings and analyzed those prompts in terms of their number, sequence, frequency, length, and content. Besides, we also analyzed students' think aloud protocols when they were thinking aloud about their prompts. We also assessed the functionality of students' apps.

To address how students talk through AI, we analyzed students' responses in their completed worksheets, transcriptions of students' presenting their responses, and think-aloud protocols when students were thinking aloud about their ideas and AI ideas.

To address how students talk about AI, we analyzed students' responses in completed worksheets, video recordings of their response presentations and think-aloud protocols.

Additionally, we adapted a draw-a-picture technique (Hwang et al., 2023) and asked students to submit AI-generated images for AI and AI in education, and the prompts to generate those images, immediately before and after the workshop: we analyzed and compared in the pre- and post-surveys students' AI-generated images and the corresponding prompts.

We present student learning evidence from the MYP design cycle chronologically: from what each student evidenced before the workshop; during the Inquiring and Analyzing and Developing Ideas stages; during the Creating the Solution and Evaluating stages; and after the workshop. We subsequently discuss these results by how students talk to, through, and about AI.

### *3.1. Student A*

#### *3.1.1. Pre-workshop*

Student A's prompts to generate a representative picture of AI and AI in education (see Figure 3) were, "Generate a representative picture of AI" and "Generate a representative picture of AI in education," respectively.

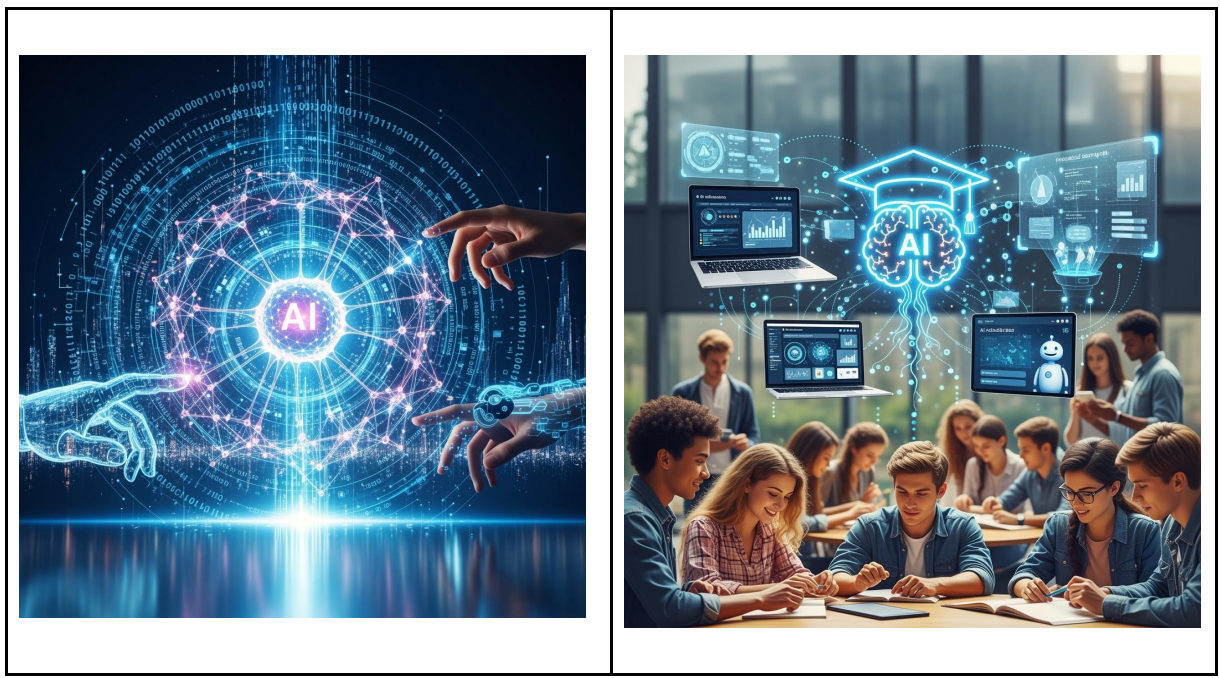

**Figure 3.** Student A's AI-generated representative pictures of AI (left) and AI in education (right)
**Alt text**: Two pictures with the left picture showing three hands touching a digital AI in the center and the right showing people gathered with computers and an AI hologram above them

#### *3.1.2. Inquiring and Analyzing and Developing Ideas*

Student A conceptualized the design problem as enabling Form 4 students to "learn to practice process writing independently." Her analysis of an existing solution identified self-regulated learning, noting that "students can set individual goals and personal writing goals and progress tracking," which work together as students "set achievable goals [and use] a simple tracking system like a checklist to monitor their progress." Her success criterion centered on user experience, stating her "design must have a friendly feedback interface, so that it could have an easy navigation for the students." She sketched her solution's chat-based interface where a student submits writing and a bot provides feedback (see Figure 4). She would name her app Your Friendly Learning Buddy.

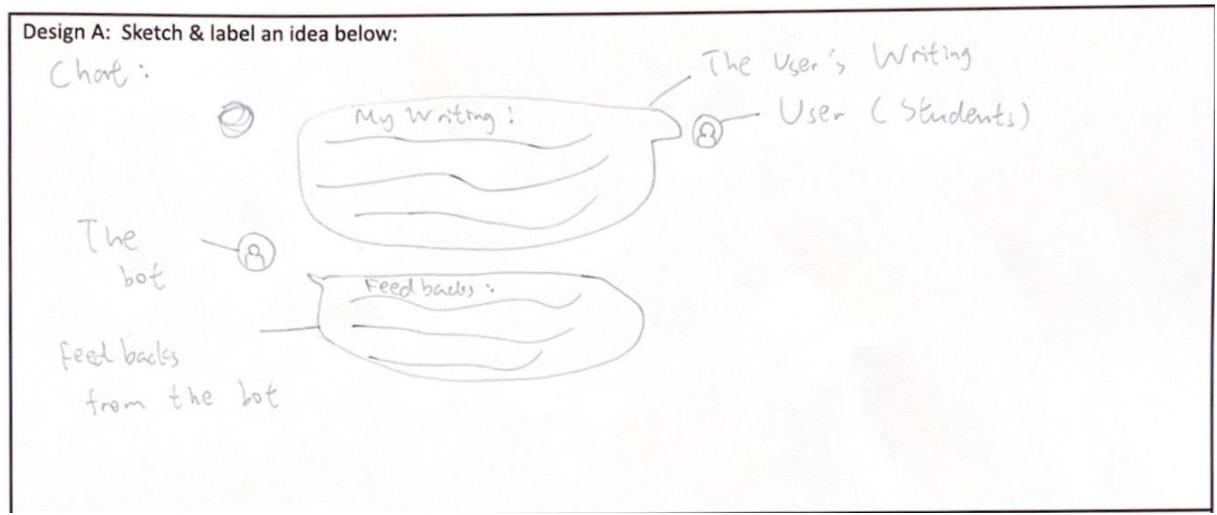

**Figure 4**. Student A's sketch and label of her app interface
**Alt text**: A drawing of an app interface with labels

Her planning was systematic. She intended 85% of the app's ideas to come from her and 15% from AI. She identified key materials such as the "HKDSE marking guidelines" and AI tools like "POE and creative and reflective prompts," with a first step to "get the HKDSE marking guidelines... which should take about 20 to 30 minutes."

*3.1.3. Creating the Solution and Evaluating*

Student A vibe coded her app on POE's App Creator. She wrote four prompts totaling 2,216 words or 554 words per prompt. Her prompts included structured context (i.e. paragraphed sections with headings) and one-shot learning, specifically, a sample composition with scores and feedback pasted into the prompt. Her strategy was to iteratively expand a single prompt, for instance, by adding sections or appending information to a section with each successive prompt. She developed her comprehensive prompt in a notes application before pasting each successive prompt into a new App Creator window in POE.

Student A's think-aloud protocol showed clear progress in completing the app. In Round 1, she reported minimal progress (1.5%) and attributed only 30% of the ideas to herself; she reported her prompt as neutral ("not good, but not bad") and her affect as "fine." By Round 3, Student A's reported completion jumped to 50%, but her perceived authorship remained at 30%. She metacognitively engaged with her prompt's structure, stating the prompt had "a lot of details." However, her affect shifted to "frustrated" because of uncertainty about the final output quality. In Round 4, Student A reported her app as 90% complete with 40% of the ideas her own. Her affect returned to a neutral "okay," and she evaluated the output as acceptable ("okay... Not that bad").

Student A's submitted app, "Your Friendly Learning Buddy," realizes her design goal of a functional interface featuring writing task prompt and composition input boxes (see Figure 5), a word count display, and a "Get Feedback" button (see Figure 6) that generates content, language and organization scores and qualitative feedback (see Figure 7).

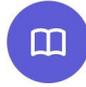

**Figure 5**. Student A's actual app interface
**Alt text**: An app interface featuring a title and two text input boxes

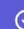

**Figure 6**. Student A's app interface with a composition pasted into the input box
**Alt text**: A composition pasted into a text input box with word count shown

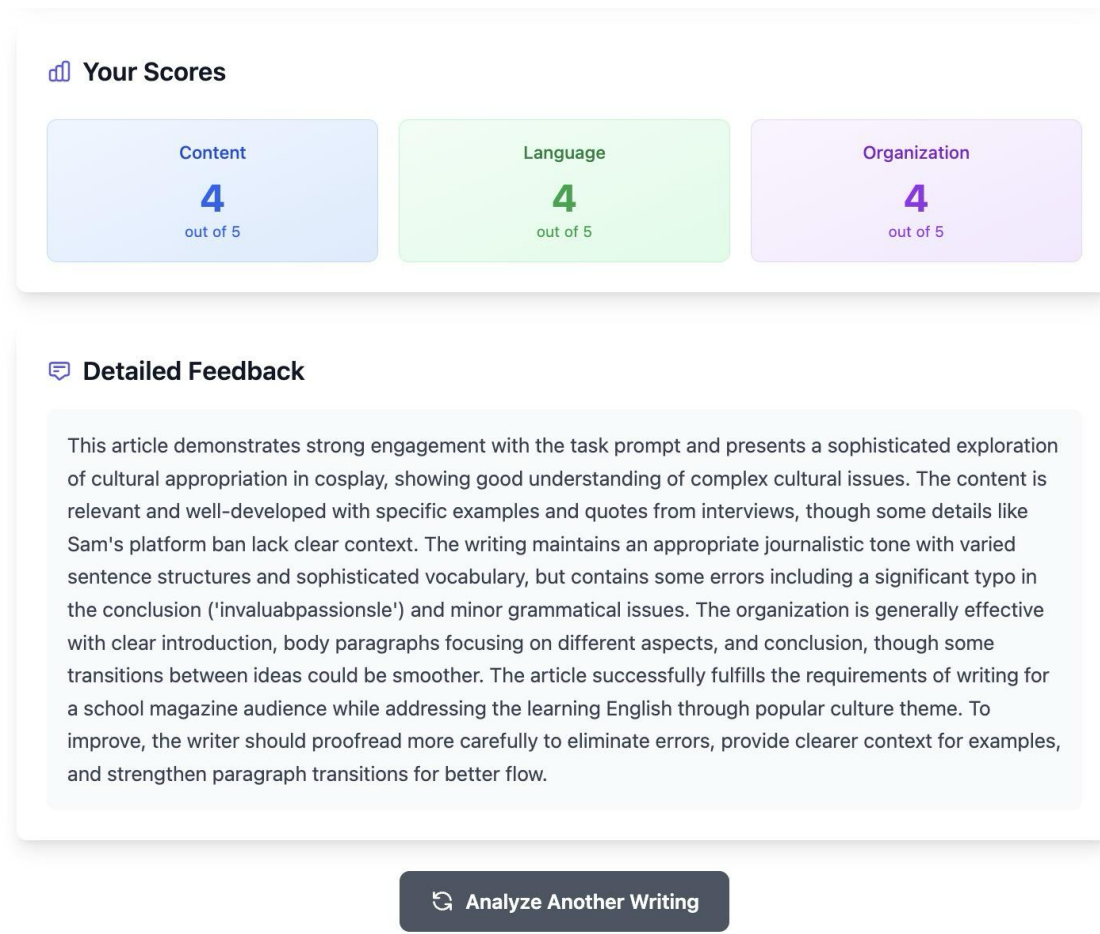

**Figure 7**. Student A's interface provides composition feedback
**Alt text**: An interface shows numerical scores and detailed qualitative feedback for a composition

      To test its functionality, Student A would employ a direct validation method, stating, "I tried to give them a random prompt and a random writing to see if the app, whether it's working as expected or not." Against her success criterion of a "friendly feedback interface," she claimed that the design was successful because the app "gives good feedbacks." For future iteration, she proposed enhancing user engagement through visual design, noting, "I would improve the color so that it's more colorful and fun to learn rather than boring colors like white and black."

*3.1.4. Post-workshop*
      Student A's prompts to generate a representative picture of AI and AI in education (see Figure 8) were, "Generate a representative picture of AI" and "Generate a representative picture of AI in education," respectively.

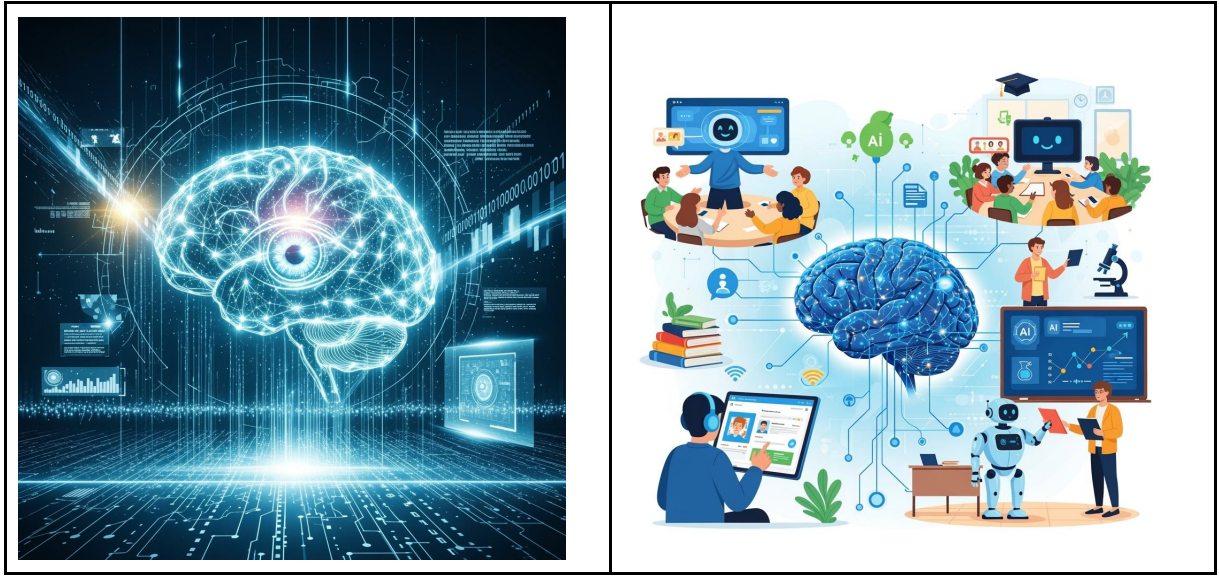

**Figure 8**. Student A's AI-generated representative pictures of AI (left) and AI in education (right)
**Alt text**: Two pictures with the left picture showing a brain hologram with an eye in the middle and the right showing human-AI interactions with a brain in the middle

### *3.2. Student G*

*3.2.1. Pre-workshop*
      Student G's prompts to generate a representative picture of AI and AI in education (see Figure 9) were, "generate the representative picture of AI" and "Generate a representative picture of AI in education," respectively.

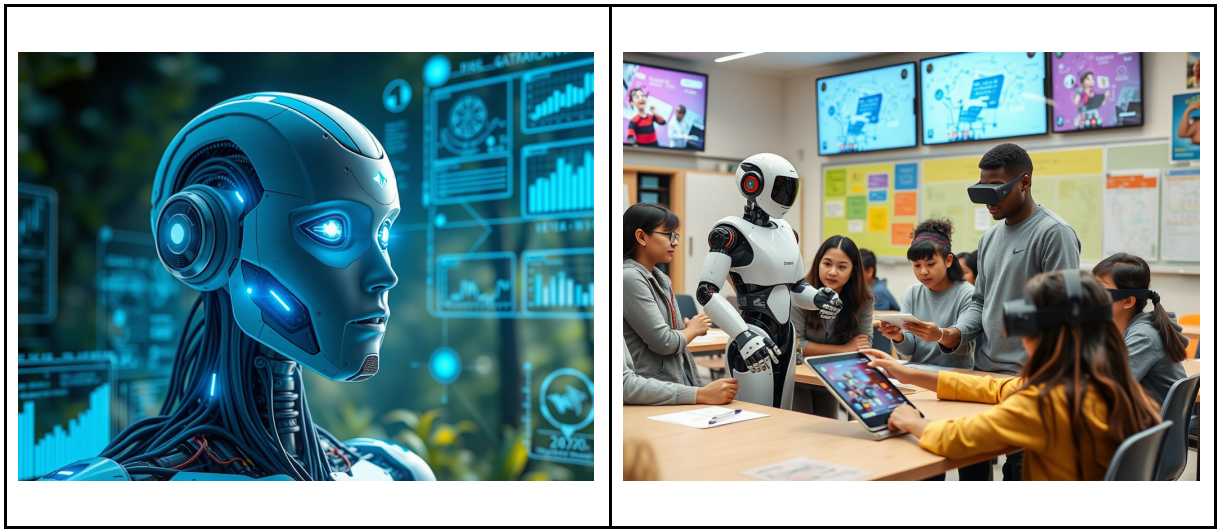

**Figure 9**. Student G's AI-generated representative pictures of AI (left) and AI in education (right)
**Alt text**: Two pictures with the left picture showing a humanoid robot and students and a robot interacting in a classroom

*3.2.2. Inquiring and Analyzing and Developing Ideas*

Student G conceptualized the design problem as enabling Form 4 students to "learn to practice process writing independently." Her research into existing solutions led her to ask Gemini-1.5-Flash search, and to identify social interaction, specifically "collaborative writing" where "students can work in small groups to complete writing tasks together" and where "interactions between students can get them to learn and they can also get feedback from one another." Her success criterion was that the "design should be engaging and not dull," as she argued a "boring design just makes people want to sleep and not use it at all." Visually, she intended the app to appear colorful, vibrant and engaging, with neon colors and a video game covering the entire screen. She sketched her solution's interface as an immersive experience where "a character is playing a video game... while the character is actually explaining the content about process writing" (see Figure 10). She would name her app Brainrot while learning!

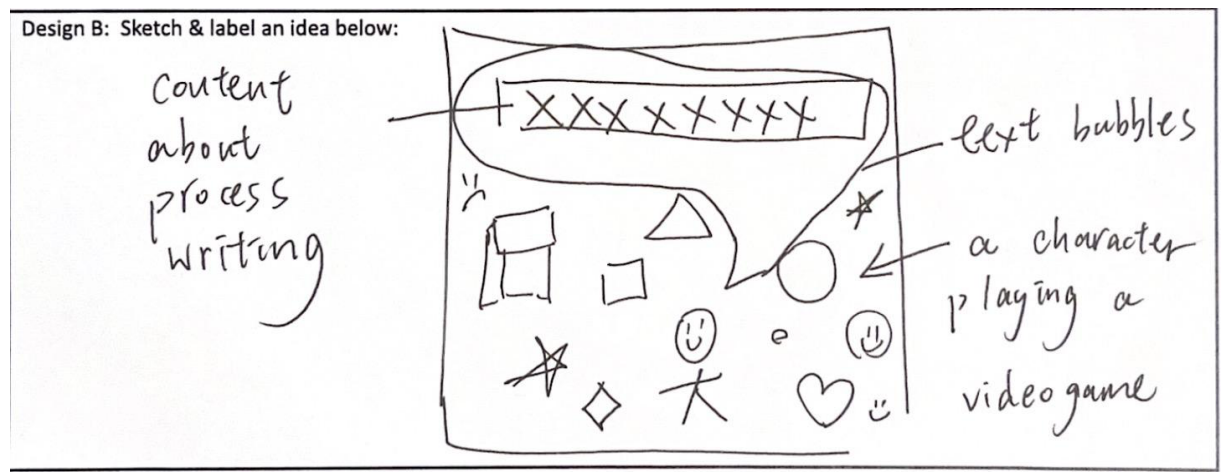

**Figure 10**. Student G's sketch and label of her app interface
**Alt text**: A drawing of an app interface with labels

Student G intended 80% of the app's ideas to come from her and 20% from AI. Her plan required materials such as "video games, like the videos," "precise sources about process writing," and "AI chatbots and AI agents because we need constant feedback," and she intended to use AI tools to produce character voices that explain process writing. Her initial step would be to "collect video game gameplay videos" estimated to take several hours.

*3.2.3. Creating the Solution and Evaluating*
Student G vibe coded her app on loveable.dev. She engineered eight prompts totaling 166 words or 21 words per prompt. Her prompts were conversational, characterized by brief requests (e.g., "Create an app with me"), reports (e.g., "I will now send you the scoring panel, feedback and writings") and questions (e.g., "Can you make it able to click in the boxes for the game play mode for users to play video games"). Her strategy appeared to iterate her app through success prompts or chain prompting. However, her vibe coding was constrained by technical difficulties: she could not access her preferred platform (Gemini), encountered errors uploading knowledge sources, and eventually reached lovable.dev's free daily message limit that stopped her progress.

Student G's think-aloud protocol reflected her difficulties. In Round 1, Student G's affect was "nervous" and "scared" as the AI could produce an undesirable output. However, she felt her prompt was "very precise" and her anxiety turned to pleasure as she saw the output, which was "really nice." In Rounds 2 and 3, Student G stalled, reporting 40% completion and 80% authorship. She noted the AI was "really slow" and her affect was "nervous" because of the time constraint. On the other hand, she expressed pride that the AI did not make something "stupid" and was surprised that AI made the app "look good." In Round 4, still at 40% completion, her affect turned to "angry" as she felt the AI's output quality had deteriorated and because she "really hope[d] it will understand" her prompts. Her affect concluded with frustration at the technical barrier of a paywall ("I feel like I'm broke"), which completely blocked her progress.

Student G's submitted app, "Writing Quest," demonstrates the student's ability to adapt while facing technical constraints. Her submitted app (see Figure 11) was a static webpage that adopted a three-section layout not in her original plan "as the AI suggested it would be simpler." Two sections, Gaming Mode and Level Progress, present gamification features. One section, HKDSE Feedback, provides HKDSE scoring dimensions and score range. At the bottom of the page is what appears to be a non-functioning button with the label, "Start Writing Quest."

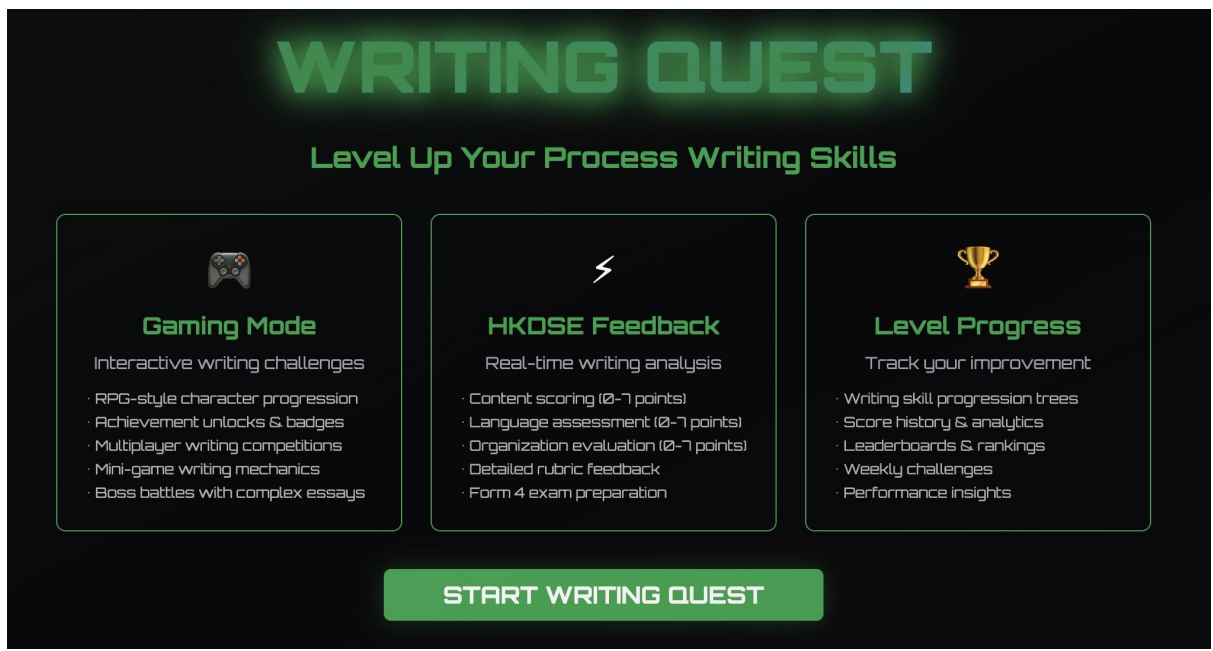

**Figure 11**. Student G's actual app interface
**Alt text**: An app interface featuring a title and three text boxes on a black background

Student G's method for testing would involve dialogue with an AI agent, explaining, "I told AI to explain how the app works, then I told the app to engage and play, and then after it plays and engaging it, I told it to share its experience." She positively evaluated the app's interface against her success criterion of creating an engaging experience, claiming she "did make it happen because the app gave a very futuristic robot video game vibe, which was not boring at all." For improvement, she focused on user onboarding, suggesting the addition of instructions and "a survey, like a small survey on how you perceive process writing" before a user lands on the main app page.

*3.2.4. Post-workshop*

Student G's prompts to generate a representative picture of AI and AI in education (see Figure 12) were, "Generate a representative picture of AI" and "Generate a representative picture of AI in education. Upload the picture," respectively.

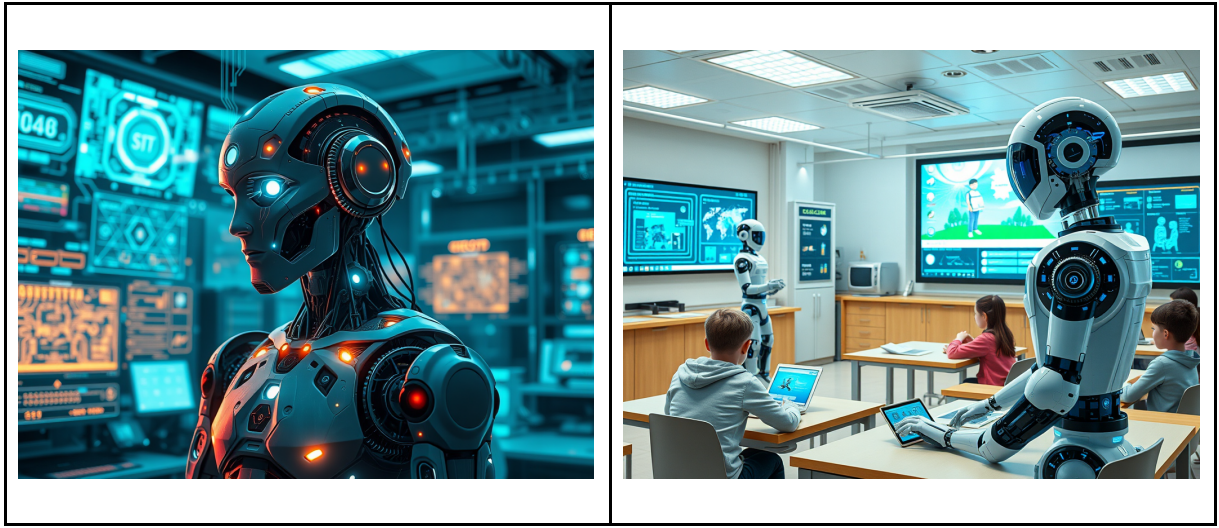

**Figure 12.** Student G's AI-generated representative pictures of AI (left) and AI in education (right)
**Alt text:** Two pictures with the left picture showing a humanoid robot and the right showing humanoid robots and students in a classroom

## 4. Discussion

This study demonstrates that in the context of vibe coding, AI can be a beneficial languaging machine (Jones, 2025), provoking EFL students' language, reflection, and self-regulation. Simultaneously, the study reveals insights into the complex interplay between how students talk to, through, and about AI. The differences in how Students A and G talk to, through, and about AI suggest how differences in these metalanguaging dimensions impact vibe coding outcomes.

Student A successfully created a functional application that cohered with her intended design. That can be attributed directly to how she *talked to AI*. She crafted extensive, structured prompts offline and showed high levels of metalinguistic awareness and metacognitive engagement. Besides, Student A did not chain prompt but required AI to execute each detailed prompt independent of her previous attempts. Although not evidenced from the draw-a-picture technique, Student A's systematic prompting suggests she views AI as a precise tool requiring detailed instructions. Curiously, although her prompt engineering strategy and mental model suggested primary authorship, Student A never attributed the majority of authorship to herself during the vibe coding task. Since Student A did intend that 85% of the app's ideas would come from her before vibe coding, it is possible that during vibe coding Student A suffered high cognitive load (Woo et al., 2024) which interfered with her perception of authorship and voice. Nonetheless, how Student A actually talked to and about AI appeared sufficient to realize her goal.

Student G showed a great gap between her creative design and the realized output. Her experience highlights another approach to *talking to AI* and its challenges. Student G talked to AI through conversational, iterative dialogue, suggesting a mental model of AI as a collaborative, creative partner. She further evidences this socially oriented way of *talking about AI* through her emotional responses. For instance, she shows pride when the AI did not make something "stupid" and frustration when it "did not understand." Curiously, although Student G's mental model of AI and her prompt engineering strategy suggest a deep, if not equal negotiation of authorship, Student G at every instance reported the majority of authorship to herself, including 80% both before and during vibe coding. Student G's emotional reactions and technical difficulties suggest she experienced a high cognitive load, which likely interfered with her perception of *talking through AI*. Ultimately, how Student G approached talking to, through, and about AI appeared misaligned and led to vibe coding performance that did not realize a functional application cohering with her intended design.

For language educators, both cases reveal opportunities to teach students to explicitly *talk to, through, and about AI.* Teachers should help students analyze how their prompting strategies (talk to) reflect AI authorship perceptions (talk through). Specifically they may lead students to critically analyze what a student's prompt engineering strategy suggests about the extent to which humans and AI share the rhetorical load and whether a product could be considered human-authored or synthetic. Besides, the students' identical, minimal prompts for generating AI images before and after the workshop suggest they either misunderstood the task or possessed a limited lexicon for articulating their imaginations of AI (talk about). If the latter, teachers can help students to develop language to describe and critically compare mental models concerning AI. Teachers should start by helping students analyze how their prompting strategies (talk to) reflect mental models. Teachers and students could also examine misalignments between students' perceptions of how they talk to, through, and about AI. Ultimately, developing language for these metalanguaging dimensions is essential for empowering students to become more purposeful designers and vibe coders.

For vibe coding educators, Student A's case shows the value of planning and precision in prompt engineering. Pedagogical activities could focus on structuring complex requests, using headings, and providing one-shot examples to scaffold students toward this more effective way to talk to AI. Student G's experience highlights the need to pair ambitious design thinking with pragmatic technical planning. Educators could start with simpler projects for students to build foundational vibe coding skills before attempting complex, multi-modal applications.

**5. Conclusion**

This innovative practice has demonstrated vibe coding's potential in EFL contexts, showing how students use natural language to design software for authentic learning scenarios. The cases illustrate that successful vibe coding is intertwined with metalanguaging practice: how students talk to, through, and about AI. Consequently, effective instruction should explicitly scaffold these three dimensions: teaching structured prompt engineering, facilitating discussions on authorship, and developing a critical vocabulary for AI.